\documentclass[journal]{IEEEtran}
\usepackage{amsmath}
\usepackage{float}
\usepackage{amsmath,amsfonts}
\usepackage{algorithmic}
\usepackage{algorithm}
\usepackage{array}
\usepackage[caption=false,font=normalsize,labelfont=sf,textfont=sf]{subfig}
\usepackage{textcomp}
\usepackage{stfloats}
\usepackage{url}
\usepackage{verbatim}
\usepackage{graphicx}
\usepackage{cite}

\usepackage{graphicx}
\usepackage{amssymb}

\usepackage{authblk}

\begin{document}
%
\title{On the Importance of Behavioral Nuances: Amplifying Non-Obvious Motor Noise Under True Empirical Considerations May Lead to Briefer Assays and Faster Classification Processes} 


\author[1]{Theodoros Bermperidis}
\author[1]{Joe Vero}
\author[1,2,3]{Elizabeth B Torres}

\affil[1]{Department of Psychology, Rutgers the State University of New Jersey, USA}
\affil[2]{Rutgers University Center for Cognitive Science}
\affil[3]{Rutgers University Center Biomedicine, Imaging and Modeling, Department of Computer Science}
   
\maketitle
   
%




%




\IEEEtitleabstractindextext{
\begin{abstract}
There is a tradeoff between attaining statistical power with large, difficult to gather data sets, and producing highly scalable assays that register brief data samples. Often, as grand-averaging techniques a priori assume normally-distributed parameters and linear, stationary processes in biorhythmic, time series data, important information is lost, averaged out as gross data. We developed an affective computing platform that enables taking brief data samples while maintaining personalized statistical power. This is achieved by combining a new data type derived from the micropeaks present in time series data registered from brief (5-second-long) face videos with recent advances in AI-driven face-grid estimation methods. By adopting geometric and nonlinear dynamical systems approaches to analyze the kinematics, especially the speed data, the new methods capture all facial micropeaks. These include as well the nuances of different affective micro expressions. We offer new ways to differentiate dynamical and geometric patterns present in autistic individuals from those found more commonly in neurotypical development.  
\end{abstract}


}


\IEEEdisplaynontitleabstractindextext

%
\IEEEpeerreviewmaketitle

\section{Introduction}

Diagnosing a disordered state of the nervous system or detecting the alteration of its normal functioning typically requires on the order of several minutes to an hour. Take for example, a speech therapist trying to discern whether a child has issues with motor planning disrupting the speech production, even when the child may seem to understand what is being asked from him. In such cases, it takes several repetitions of various facial motions to perhaps conclude the presence of apraxia impeding the motor planning, sequencing, coordination and control of facial muscle units. Likewise, a diagnosis of impairments in social interactions and communication, also known as autism spectrum disorders, may take between 45 minutes to an hour of trying to engage the child with a clinician that provides social presses to measure the overtures of the child in response to various similarly predefined social and emotional prompts. Trial by trial, the observer searches for key visible features that can be reliably detected by the naked eye, while filtering out non-obvious ones that are rendered as noise or gross data. But what if trapped in those behavioral nuances are important signals that reveal much faster, in very brief assays, the departures from typical functions of the nervous system?

Exclusive reliance on obvious features of phenomena extends beyond the realm of observation to the context of data analysis. Often in the statistical analysis of behavioral data from motor actions, the researcher adopts parametric models that rely on theoretical assumptions of linearity, stationarity and normality of the data under consideration. This often leads to neglecting spontaneous activity hidden interspersed between more visibly reliable segments of behavior, i.e., what we call behavioral nuances or gross data and discard in grand averages across epochs that an observer may (sometimes even arbitrarily) preset as the relevant phenomena (Figure 1). 

It has been our finding for nearly two decades now, that biorhythmic data in the form of time series harnessed by biosensors is best described by nonlinear processes that are non-stationary and poorly fit by the normal distribution  \cite{1} \cite{2},  \cite{3}. From electroencephalographic data to electrocardiogram and naturalistic, complex bodily movements, we have systematically found in neonates  \cite{4}, children \cite{3} and adults \cite{5} that the biorhythmic data registered from the nervous systems' responses to a plethora of sensory stimuli, needs to be re-examined through a different empirical lens, one that must include behavioral nuances as well as the visibly reliable segments \cite{2}. Along this journey, we have found that the nuances treated as gross data not only give us important signals but also can reduce the time of our assays to probe the nervous systems’ functions  \cite{6}, \cite{7}. We do so by amplifying features of the motor noise that reveal different pathological states along with their evolution over time. 
We present here new ways to re-examine time series data from human biorhythms of facial micromotions harnessed through video recordings that leverage recent AI techniques (e.g., OpenFace \cite{8}) estimating the facial grid. Using our approach, tracking the facial micromotions for merely a few seconds is sufficient to discern differences across a random draw of the population and use a machine learning classifiers to automatically separate autistic individuals from neurotypical controls. We propose adopting such methods of non-obvious nuances, to avoid taxing the person with long diagnostic assays, while also promoting new ways to scale our research beyond the confines of highly controlled, lengthy experiments in our labs. The work presented here is an invitation to embrace true empirical features of our behavioral data and discover new dimensions of information in the behavioral nuances, rather than a priori determining what we are already biased to find under rigid theoretical constraints.

\section{DATA AND METHODS}

\subsection{Facial kinematic data}

This study used a research app to register, in brief intervals of 5 seconds, facial expressions during each of a practice round, at rest, smiling, and making a surprised face (spanning 20 seconds in total). The participants received instructions through the app, and an emoji projecting the desired face (neutral, smiling or surprised) was used to prompt the desired expression. The app recorded the videos under natural conditions (at school, home, social events, and at conferences). 

The participants comprised a cohort of 124 individuals, 75 neurotypicals (NT) and 49 on the autism spectrum (ASD). The ASD participants included 20 who needed low support (LS ASD) and 29 who needed high support (HS ASD). IRB approval was obtained for all participants. Although the former had subtle sensory and motor issues, the latter had visible sensory-motor differences and minimal-to-none articulated vocal sounds. The study aimed at characterizing the stochastic properties of the moment-to-moment variations in these empirical face data. The analyses seek to uncover parameters that enable a compact representation of the nonlinear dynamics of the facial motions. We present digital biometrics capable of characterizing normative ranges of various parameters, along with departures from such ranges in participants with different neuro-developmental trajectories, such as those with a diagnosis of autism spectrum disorders.

The video data was used as input to the OpenFace platform to generate a grid of 68 landmarks along with the activation of 17 muscle Action Units \cite{8}.

The 2-D positional pixel coordinates of the landmarks are realized within a virtual rectangular grid centered around the participants' face. These positional pixel trajectories were smoothed and differentiated using splines to obtain the velocity and acceleration fields. These in turn provided the scalar speed and acceleration time series, which we obtained using the Euclidean norm. We applied z-normalization to the positional time series data to account for differences in distance to the camera and designed visualization methods to immediately see the outcome of the tasks. Figure 1A on the left shows the normalized facial traces of the pixel positional trajectories of each of the 68 grid points. These are from a representative NT at rest. The data from a representative ASD participant, also at rest, is shown on the right. The positional trajectories in both cases show the amount of involuntary micromotions that spontaneously occur at rest (i.e., without instruction), for each representative participant. Figure 1B shows a visualization example to capture the amount of motion at rest. By making the size of the marker on each grid point proportional to the length of the path traveled during the 5 seconds of this resting state, we can appreciate the amount of involuntary motions that the person has. One can also immediately appreciate the profound differences in energy consumption and effort that the brain will have to exert in order to maintain a steady gaze that can bring a reliable image to the retina. Then, further sensory-to-motor transformations will have to take place to coordinate the hand and the eyes to, for example, make a pointing gesture to a visual target while attempting to communicate a decision.

\begin{figure}[ht]
\centering
\includegraphics[width=0.4\textwidth]{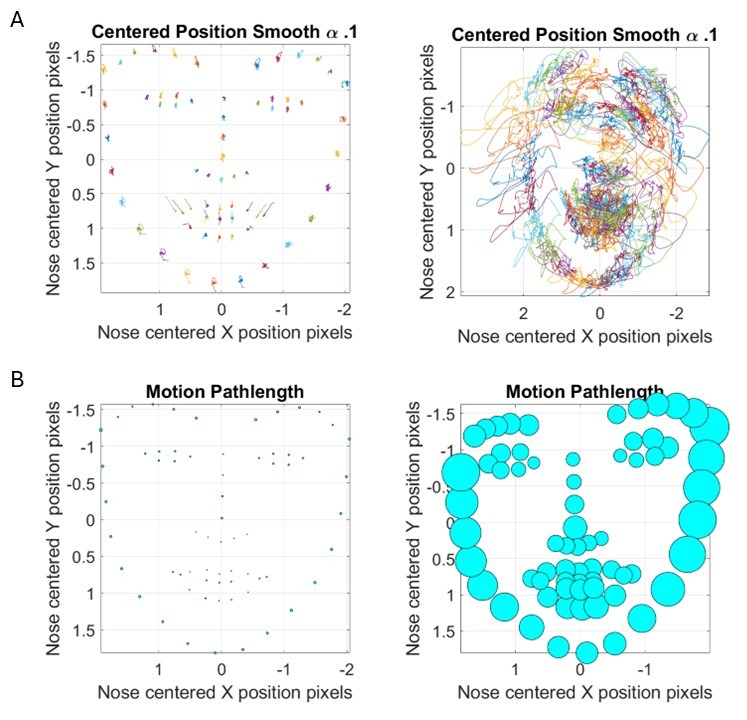}
\caption{Facial micromotions at rest for two representative participants (A) Neurotypical vs. ASD. (B) Representation of the path length from the motion excursion traveled at each point of the grid to immediately visualize the outcome of the task and to assess at rest, the amount of involuntary motions (or the amount of volitional control) that the person has when asked to keep still.  
  }
\label{fig:Panel 1}
\end{figure}

We adopt the methodology developed by our Sensory Motor Integration lab \cite{3}, to characterize human biorhythmic data samples registered from different functional layers of the nervous systems. These consist of time-series data recorded from participants throughout the human lifespan (from neonates to the elderly). The recordings include electroencephalogram \cite{9}, auditory brainstem responses \cite{10}, electrocardiogram \cite{11}, bodily acceleration \cite{12} \cite{13} \cite{6}, rotational data from gyroscopes\cite{14}, and bodily kinematics during various functional activities such as resting \cite{15}, pointing naturally \cite{3} \cite{16}, walking \cite{17}, dancing \cite{18}, and playing sports \cite{2} \cite{19}, among others. The details of this empirically informed body of work can be found in the literature; here we will provide a brief overview of the methodology.

Using the linear speed or linear acceleration data, we obtain the peaks of the waveform and extract features such as the amplitude, the prominences, the widths and the inter-peak interval times. These peaks are gathered in a frequency histogram, and using maximum likelihood estimation methods (MLE), we obtain the best-fitting continuous family of probability distributions. This is the Gamma family with shape and scale parameters. The first moment (the Gamma mean) is then obtained, and the absolute deviations from the mean are used as the new time series to consider, while preserving the time indexes of the original peaks. These positive deviations are then subject to a normalization that scales out allometric effects \cite{20} due to disparities in facial lengths between participants, a factor that influences motion speed. The normalized peaks form the waveform that we have coined micropeaks spikes (MMS) or micropeaks. Such micropeaks are also best fit by the continuous Gamma family of probability distributions. 

Our research has found that across the human lifespan, and for goal-directed voluntary movements, the scatter of points spanned along the log shape and the log scale axes, on the Gamma parameter plane, follows a tight linear fit. Knowing one parameter can thus accurately infer the other. This empirical finding, paired with the fact that in the continuous Gamma distribution case the noise-to-signal ratio (NSR) defined by the variance divided by the mean, is the scale value (dispersion); let us then use this parameter to summarize several important features of the phenomena under consideration. Furthermore, pairing this parameter with others can lead to useful parameter space identification and new classification schemes \cite{3}. Formally:

\begin{equation}
 v_{absolute}=\vert v-E[v] \vert \Rightarrow s= \frac{v_{absolute}^{peak}}{v_{absolute}^{peak}+v_{absolute}^{local average}} 
\end{equation}

The mean is subtracted from the amplitude values across speed series, and then these peaks are normalized by the local average. Peaks at mean value are set to 0, that is, they are the Gamma mean of the original peak series. Then, the random variable $s$ follows a gamma distribution:

\begin{equation}
s\sim \Gamma(k,\theta)
\end{equation}

where our lab has discovered \cite{3} that the log shape and the log scale empirically satisfy a power law of the form:

\begin{equation}
k=a*log(\theta)+b, \frac{\sigma^2}{\mu} = \theta
\end{equation}

As the motor system matures, the micropeaks of the movement speed exhibit lower $NSR$ (scale) and higher shape values. When typical development is disrupted or delayed (as in the case of ASD), this is not the case \cite{3}, the $NSR$ in fact, increases with age \cite{21}

\begin{figure}[ht]
\centering
\includegraphics[width=0.4\textwidth]{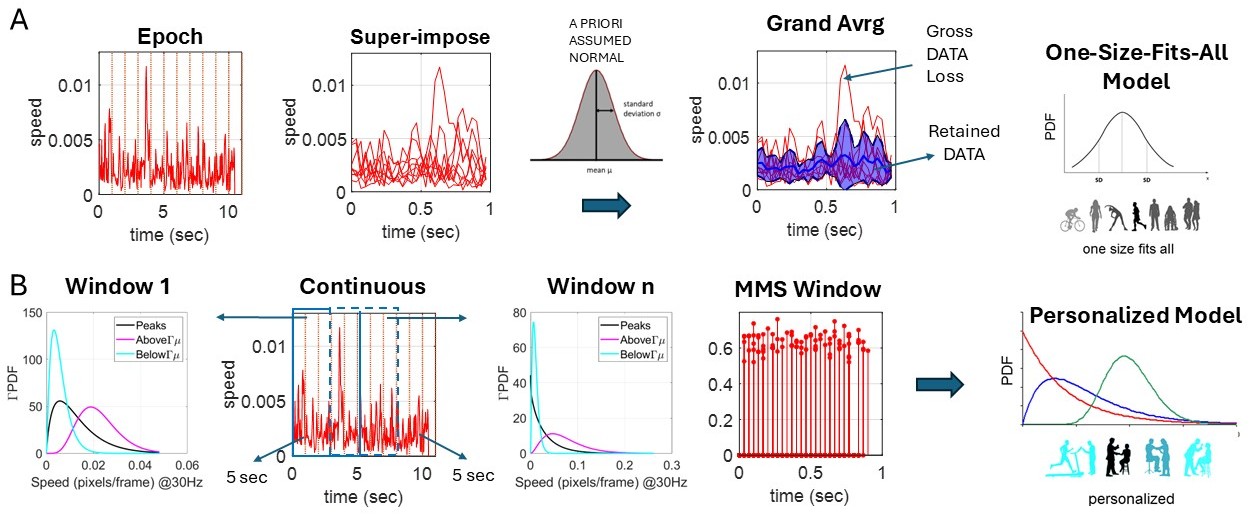}
\caption{Traditional one-size-fits-all model vs. proposed personalized approach. (A)  In a typical scenario, a z-normalized speed time series is split into non-overlapping epochs (e.g., 1 second-long windows). The marginal behavior of the stochastic process at time $t$ is “a priori” assumed to follow a normal distribution with mean $\mu_t$ and standard deviation $\sigma_t$. The grand average over the stacked up windows under this theoretical assumption then discards the gross data, outside the standard deviation bounds. (B) The proposed model considers instead the continuous, moment-to-moment micropeaks in the speed time-series. In 5-second-long windows (at 30Hz sampling resolution in this case) we empirically estimate the best continuous distribution family in an MLE sense (see text), obtaining the first moment, and the deviations from it, and then normalizing the peaks by their local average. By scaling out in this way allometric effects due to anatomical disparities across participants, we standardize the waveform to unitless 0-1 valued time-series and can properly compare individual differences in a truly personalized way. The micropeaks thus obtained provide the empirical histograms for each window. Each window (with 50 percent overlap in this case) is best fit in an MLE sense, by the continuous Gamma family of probability distribution functions. Shown here for two windows are the estimated Gamma PDFs from the raw data peaks' amplitudes used to obtain the empirical mean of the window along with the empirical distributions of the micropeaks deviating above and below the empirical mean. This approach captures the non-stationary nature of the time-series data and allows for an objective and personalized characterization of motor/ behavioral repertoires.  
  }
\label{fig:Panel 2}
\end{figure}

Figure 2 compares the traditional approach of grand averages of biorhythmic time series under the a priori assumption of normality vs. our personalized approach using micropeaks. Instead of averaging out gross data, micropeaks (MMS) capture statistical nuances that enable empirical characterization of behavioral data. This provides us with personalized biometrics, as opposed to a one-size-fits-all model.  

\subsection{Tracking Facial Synergies in the Motor Code}
Facial synergies occur spontaneously and largely under awareness, as we engage in social and emotional exchanges and as we communicate, eat (masticate and swallow) or even rest the muscles of the face (in the cases where involuntary motions engage the facial muscles). To assess the phenomena involving the closed-loop between the face grids under study and the underlying action units characterized by the field, we introduce two hierarchical levels of organization in the model that we propose. The first refers to a macro-level and the second to a micro-level of synergistic behaviors.  

\begin{itemize}

\item Macro-level: A causal level of muscle activations, described by the 17 action unit time series, is our first level of analysis. Muscle units are activated to contract muscle fibers across the face. Each muscle group exerts force across a wider area of the face; e.g., in the case of the eyebrow raiser, it pulls the entire eyebrow upward. We consider this to be a level of muscle dynamics that causes changes in a second level below.

\item Micro-level: The combined AU activations at the macro-level create local motion on the surface of the skin. Moment-to-moment, the grid points describing the changes of position over time produce the time series of speed values used to obtain the micropeaks (MMS) described in Figure 2B. This micro-level of facial landmark kinematics is then used as a proxy of the quality of motor feedback (kinesthetic reafference) that the system experiences from moment to moment. If this code is random noise, the feedback is poor. If this code is predictive, with a high signal-to-noise ratio, the feedback is of high quality. As such, this framework provides the grounds for interpretation of the digital data, particularly when we characterize the normative ranges of neurotypical controls and measure the departures from these normative ranges in autistic participants.
\end{itemize}

To further facilitate our interpretation while considering multiple neurological conditions such as HS ASD with a diagnosis of apraxia (such as those HS ASD in this work), we rely on the zoning of the trigeminal nerves innervating the ophtalmic, maxillary and mandibular areas, denoted V1, V2 and V3, respectively, in Figure 3.

We can quantify the level of synergies between landmark speeds or Action Unit activations by computing covariance matrices that capture correlations between time series across the face. Then, using concepts and methods from the theory of positive definite matrices PDMs, we develop a novel and biologically intuitive concept of "synergy" that relies on the geometric characterization of spaces of (PDMs) \cite{22}.

\subsubsection{The covariance matrix as a geometric description of facial synergies}

Assume a collection $\boldsymbol{X}=(X_{1},X_{2},...,X_{n})$ of random variables. The covariance matrix $K_{\boldsymbol{XX}}$ is the matrix with entries in $(i,j)$:

\begin{equation}
K_{X_{i},X_{j}}=cov(X_i,X_j)
\end{equation}

Because covariance is a symmetric operator in the vector space of random variables, the covariance matrix is symmetric. Moreover, it can be shown that it is also positive semidefinite \cite{22}. Generally, a matrix $M$ is positive semidefinite if: 

\begin{equation}
x^{T}Mx \ge 0
\end{equation}

for every real vector $x$. If it is strictly larger than zero, then the matrix is positive-definite. In this case, the columns of the matrix are linearly independent from each other, i.e. no non-trivial linear combination of the column vectors is zero. Positive definite matrices form a vector space, which, when endowed with the appropriate metric, is a smooth, continuous and differentiable Riemannian manifold.  \cite{23} \cite{24}

If a set of random variables is independent in the above sense, then we can associate with that set a positive-definite covariance matrix that captures correlations between the variables and lies somewhere on that manifold. We will see that this will allow for a compact description of facial synergies in terms of shortest distances (geodesic distances) in the multivariate space of landmarks' series of micropeaks. 

\subsubsection{Log-Euclidean distance between positive definite matrices}

Generally, if $M$ is a manifold, we can associate at each point $p \in M$ a vector space $T_{p}M$, called the tangent space to the manifold at that point. In the case of objects such as a sphere in a 3-D space, this is what we intuitively and formally know as the tangent vectors at a point of the surface. 

If we also endow this vector space with a Riemannian metric, i.e., a positive definite inner product with "smooth" properties $g_p: T_{p}M \times T_{p}M \rightarrow R$, the manifold is called "Riemannian". The inner product induces the norm $v_p=\sqrt{g_p(v,v)}$. In such a manifold, we can define continuously differentiable curves and compute distance along paths. 

For a continuously differentiable curve $\gamma:[a,b] \rightarrow M$, we can compute its length:

\begin{equation}
L(\gamma)=\int_{a}^{b} \sqrt{g_{\gamma(t)}(\dot{\gamma}(t)\dot{\gamma}(t))} \,dt
\end{equation}

The path with the shortest (geodesic) distance between two points in the manifold is called a "geodesic curve". Moreover, at each point in the manifold, a unique geodesic emanates in the direction of a specific tangent vector. \cite{25}

It has been shown that the space of positive definite matrices (in this case of covariance matrices) forms a Riemannian manifold under the appropriate metric. In this project, we follow the log-Euclidean metric, introduced by Arsigny  et. al \cite{26}, which has been shown to have advantages over other metrics \cite{27}. For a comprehensive review on the different metrics as well as an in-depth presentation of the prerequisite math to understand this framework, see \cite{23}. 

Under the log-Euclidean metric, the geodesic distance between two positive definite matrices $P_{1}$ and $P_{2}$ is shown to be equal to the log-Euclidean distance:

\begin{equation}
d(P_{1},P_{2})= \| log(P_{1})-log(P_{2}) \|_{F}
\end{equation}

Here, $\| . \|_{F}$, is the Frobenius norm of a matrix \cite{28} and the logarithm is the matrix logarithm, which is the inverse function of the matrix exponential.

\begin{equation}
e^{A} = \sum_{n-0}^{\infty} \frac{A^{n}}{n!}
\end{equation}

The logarithm is unique if and only if the matrix has real eigenvalues, which is the case for positive definite matrices. \cite{29} For computing the log-Euclidean distance, we used the SPDtoolbox \cite{51}, for the associated paper, see \cite{50}.

\subsubsection{Facial synergies as a shortest path distance from independence}

Consider the set of facial landmarks $i=1,...,68$ with their corresponding normalized speed micropeaks $v_{i}(t)$, as functions of time. The covariance matrix:

\begin{equation}
K_{v_{i},v_{j}}=cov(v_i(t),v_j(t))
\end{equation}

has as diagonal elements the speed variances of each landmark, and the non-diagonal elements are the correlations between pairs of landmark micropeaks processes. 

If there were hypothetically zero synergies between any two landmarks, then we can assume that the respective processes would be statistically independent and thus the pairwise covariances equal to zero. We can express this hypothetical scenario with a covariance matrix $K_{independent}$, whose diagonal elements are the variances but all other elements are equal to zero. Formally:

\begin{equation}
K_{independent}=diag[cov(v_i(t)),cov(v_i(t)]
\end{equation}

We then define a new biometric, called "Synergies", as the log-Euclidean distance of the covariance matrix from the matrix $K_{independent}$. This is the relative departure of a network of landmark speed micropeaks from a theoretical state of zero dependencies:

\begin{equation}
D_{synergies}=\| log K-log K_{independent} \|_{F}
\end{equation}

Figure 3 shows the computational pipeline, from normalized positional data to the synergies metric, using as an example the landmarks of the V3 area, representative data comprising an ASD and an NT participant.

\begin{figure}[ht]
\centering
\includegraphics[width=0.4\textwidth]{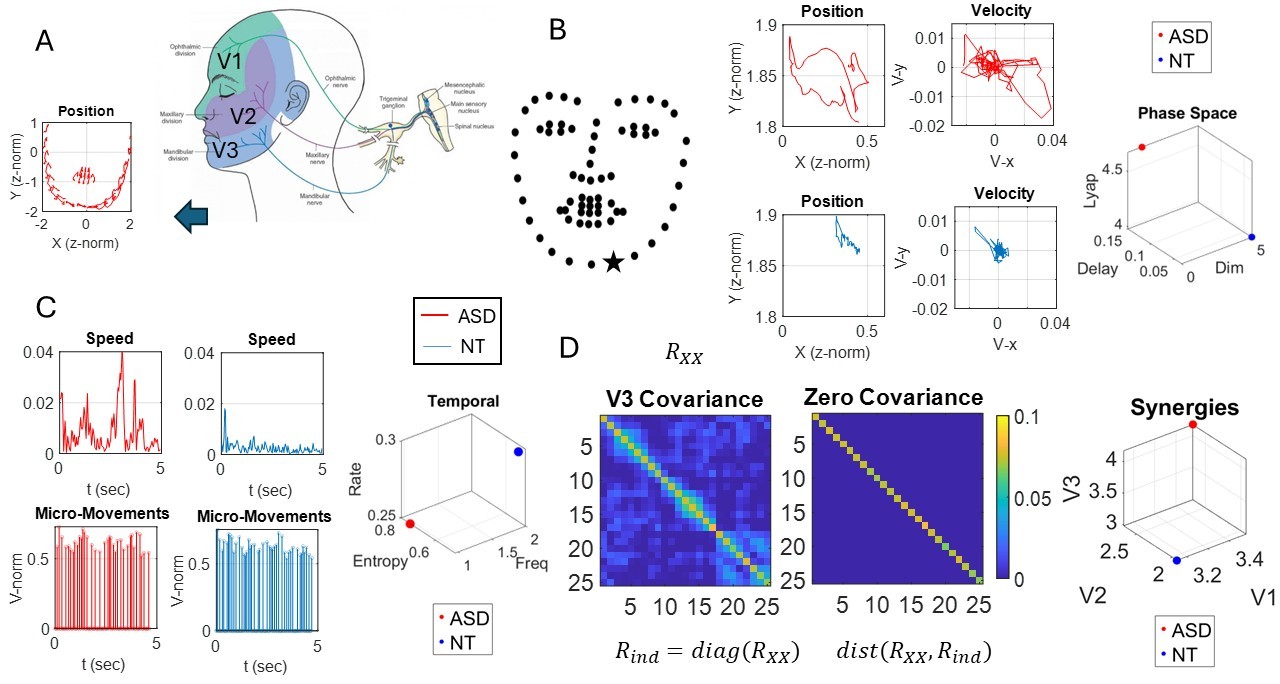}
\caption{Proposed pipeline for the characterization of facial landmark kinematic data combining dynamical systems theory and the geometry of positive definite covariance matrices to quantify facial synergies. (A) Facial landmarks obtained from the open-source software OpenFace are partitioned into three distinct areas as innervated by the trigeminal nerves: lower face (mandibular), mid face (maxillary) and upper face (opthalmic). (B) Positional trajectory from a landmark in the mandibular area and corresponding velocity field (one representative ASD and one NT participant). Corresponding parameter space spanned by dimension, delay, and Lyapunov exponent along the x, y, and z axis respectively, with each point localizing one representative participant. (C) Derived temporal speed profiles and micropeaks from each representative case. Another parameter space (frequency, entropy, rate) along the x, y and z axis respectively. (D) Covariance matrix capturing correlations between the landmarks. If we omit non-diagonal elements from the covariance matrix, we obtain the “zero covariance” matrix which only contains diagonal elements of gamma variances and represents an idealized state of zero landmark correlations. Parameter space defining “synergies” as points along the face regions V1-V2-V3 denoting the log-Euclidean distance between these two matrices in the space of positive definite matrices, i.e., how far away a particular state is from a state of no synergies. }
\label{fig:Panel 3}
\end{figure}

\subsection{Assessing Features of the Underlying Dynamical System of Individual Landmarks}

\subsubsection{phase-space Parameter Identification}

Besides the synergies, we seek to characterize the local temporal features of micro expressions derived from facial gestures. Specifically, we focus on the velocity of each landmark and aim to derive features from its reconstructed phase-space. 

We used a method proposed by Wallot et al. \cite{30}, which generalized the phase-space reconstruction from univariate to multivariate random processes. For more information on the concept of phase-space reconstruction, we refer the reader to the appropriate literature \cite{31}. We just briefly present the basic concepts and formulae for the univariate case. 

Let $x(t)$ be an observed time series. In real life applications, a time-dependent observation is often stochastic; we can, however, make the assumption that there exists an underlying dynamical system in a higher-dimensional space whose sampling period (known as "embedding delay") is larger than that of $x$, namely the Reconstructed Space. Then, the random observation $x$ is a projection of a trajectory in this space manifold to our measurement device. If we denote the reconstruction dimension as $m$ and the time delay as $\tau$, then the reconstructed variable has the form:

\begin{equation}
\boldsymbol{S_{t}} = (x_{i},x_{i+\tau},...,x_{i+\tau m})
\end{equation}

The optimal time delay $\tau$ is chosen to minimize the average mutual information between consecutive samples:

\begin{equation}
\tau_{optimal} = {argmin}_\tau \sum_{t}^{}p(x_t,x_{t+\tau})log(\frac{p(x_t,x_{t+\tau})}{p(x_t)p(x_{t+\tau})})
\end{equation}

We can view mutual information between consecutive samples in phase-space as the amount of redundant information, that is, information that is already present in the previous samples. Thus, minimizing mutual information is equivalent to maximizing the novelty of a new observation\cite{32}.

Once we have a reliable estimate of $\tau$, we can approximate $m$, using the False-Nearest-Neighbors (FNNs) technique \cite{32}. Briefly, if $m$ is the "true" dimension of the reconstructed system, points that are neighbors in $m$ dimensions will also be neighbors in $m+1$ dimensions.

We computed the reconstructed dimensions and delays for the multivariate velocity time series of each landmark. Note that the method that we used outputs the embedding dimension of a single component of the velocity vector, the dimensionality of the entire system would be equal to $3m$, since we have three coordinates in total. Throughout this paper, by choice, we will report $m$ and not $3m$.

\subsubsection{A Dynamical Notion of Predictability}

Certain real biological systems can be unpredictable and sensitive to initial conditions, that is, chaotic \cite{33}. Small deviations in the present trajectory of a system could lead to larger deviations from the expected (predicted) future trajectory. Alternatively, for certain systems, initial perturbations do not affect future outcomes much, and different trajectories may converge to the same area in phase-space. Quantifying this dynamical notion of predictability and applying it to kinematic data could provide us with a way of quantifying the stability of the motor system underlying facial micro expressions.

Dynamical system stability can be assessed by deriving constants known as the "Lyapunov exponents". Here we briefly present the basic concepts and how they can be applied to facial landmark velocity data. We will present the general framework \cite{34,52} and the variation that we used in this study, developed by \cite{35} and available in Matlab \cite{36}. Let:

\begin{equation}
\boldsymbol{S_{t}} = (x_{i},x_{i+\tau},...,x_{i+\tau m})
\end{equation}

be the reconstructed phase-space for one of the velocity components. We formulate a typical trajectory as follows:

\begin{equation}
{S_0,S_1,S_2,...,S_t,...}
\end{equation} 

A deterministic dynamical system will be equipped with the map:

\begin{equation}
S_{t+1}=F(S_t)
\end{equation}

Consider a small perturbation $u_t$ about this trajectory:

\begin{equation}
S_{t+1}+u_{t+1}=F(S_t+u_t)
\end{equation}

Assuming that the function $F(.)$ is smooth and differentiable, it can be approximated through its Taylor expansion around the origin $S_t$. Ignoring higher-order terms, we eventually arrive at the expression:

\begin{equation}
u_{t+1}=DF(S_t)u_t
\end{equation}

where $DF(S_t)$ is the matrix of derivatives of the components $i$ of $F(S_t)$ with respect to the coordinates $j$ of $S_t$, in other words, its Jacobian matrix:

\begin{equation}
DF(S_t)=A_{ij}=(\frac{\partial F_i}{\partial S_j}) \Bigr\rvert_{S = S_t} 
\end{equation}

We have arrived at an expression that shows how the error between the initial trajectory and the perturbed one evolves over time. If we sequentially apply the mapping (18) to $n$ consecutive samples, starting from $S_0$ we get the following:

\begin{equation}
u_n=A_{n}A_{n-1}...A_1 u_0
\end{equation}

We will refer to the product of the $n$ consecutive matrices as $DF^{n}(S_0)$ and we are interested in the behavior of the $l^2$-norm of the error:

\begin{equation}
\| u_n \|=u_{n}^\intercal u_n =u_{0}^\intercal DF^{n}(S_0)^\intercal DF^{n}(S_0) u_0 
\end{equation}

The matrix $DF^{n}(S_0)^\intercal DF^{n}(S_0)$  is positive, real and symmetric; therefore, it has real eigenvectors and real nonnegative eigenvalues. We denote by $s_{ni}$ and $e_{ni}$ the i-th eigenvalue and eigenvector of the matrix $DF^{n}(S_0)$ and consider an initial perturbation in the direction of this eigenvector $u_0=a_0 e_{ni}$. It can be easily shown that the i-th eigenvalue of $DF^{n}(S_0)^\intercal DF^{n}(S_0)$ is equal to $s_{ni}^2$.

Equipped with these definitions, we can show for the relative squared error at time $n$:

\begin{equation}
\frac{\| u_n \| ^ 2}{\| u_0 \| ^ 2} = s_{ni}^2
\end{equation}

We are interested in the exponentially converging or diverging behavior of this relative perturbation quantity over $n$ samples; we are looking for a constant $\lambda_i$ such that:

\begin{equation}
e^{\lambda_i n}=\frac{\| u_n \| }{ \| u_0 \|}
\end{equation}

For $n \rightarrow \infty$, one can prove that: 

\begin{equation}
\lambda_i =  \lim_{n\to\infty} \frac{1}{n} ln \| DF^{n}(S_0) u_0 \|
\end{equation}

These are the Lyapunov exponents, which when ordered in decreasing order are known as the Lyapunov spectrum. When the largest Lyapunov exponent is negative, it means that the dynamical system is stable and neighboring trajectories converge (loosely speaking to what is known in the theory of chaotic systems as an attractor); when it is positive, it shows instability and high sensitivity to initial conditions. 

The method we follow in this paper is based on the work of \cite{37} and is available in Matlab. Calculate a single Lyapunov exponent for the entire reconstructed phase-space by first computing the local Lyapunov exponent at a point $t$:

\begin{equation}
\lambda_t = \frac{1}{K_{max}-K_{min}+1} \sum_{K=K_{min}}^{K_{max}} \frac{1}{K} ln \frac{\| S_{t+K} - S_{t^{*}+K} \| }{ \| S_{t} - S_{t^{*}} \|}
\end{equation}

where the range $[K_{min} K_{max}]$ is the expansion range around the local sample and $t^{*}$ is chosen so that $S_{t^{*}}$ is the nearest neighbor of $S_{t}$, with the constraint that $\lvert t-t^{*} \rvert \leq \frac{1}{f_{avg}}$, where $f_{avg}$ is the average frequency of the signal (computed using Fourier analysis). Essentially, the ratio $ \frac{\| S_{t+K} - S_{t^{*}+K} \| }{ \| S_{t} - S_{t^{*}} \|}$ is an efficient approximation of the relative perturbation $\frac{\| u_n \| }{ \| u_0 \|}$. 

Finally, polynomial regression is used to compute an average Lyapunov exponent from the local estimates. This method is efficient for small datasets like ours. 

\subsection{Temporal Information Theoretic Measures}

In sections B and C, we proposed biometrics that track pairwise interactions between landmarks (geometric approach) as well as their dynamical behavior from a nonlinear analysis standpoint. We now complete our analysis by considering the information capacity of the motor code that we can measure by sampling facial micro expressions. To do this, we apply the MMSs framework described in Figure 1B to continuous streams of speed data. The sequences of micropeaks (deviations away from the empirical mean of all peaks) and pauses (states of average activity) are analyzed while preserving the time stamps at which they occurred \cite{3}; in other words, a temporal code. 

The simplest quantity that we can measure is the average rate, the number of peaks per unit of time.

\begin{equation}
R=\frac{\# peaks}{\# peaks+\# pauses}
\end{equation}

Consider now $N$ consecutive micropeaks samples: 

\begin{equation}
{M_1,M_2,...,M_N}
\end{equation}

If the process is stationary, from the ergodic principle \cite{37}, we can treat each set of $N$ samples as a realization drawn from a multivariate probability space.

\begin{equation}
[M_t,...,M_{t+N}] \sim (\mathbb{R}^{+}\cup {0},B(\mathbb{R}^{+}\cup {0}),P[M_1,M_2,...,M_N])
\end{equation}

where $\mathbb{R}^{+}\cup {0}$ is the sample space of micropeaks (MMS), $B(\mathbb{R}^{+}\cup {0})$ the associated Borel set, and $P[M_1,M_2,...,M_N]$ the joint probability distribution. 
Then, if $H(M_1,M_2,...,M_N)$ is the Shannon entropy \cite{38} of this random variable, the quantity:

\begin{equation}
H = \lim_{N\to\infty} \frac{1}{N} H(M_1,M_2,...,M_N)
\end{equation}

converges, is known as the Entropy Rate and measures the information uncertainty of the stochastic process or, equivalently, its degree of unpredictability from an information theoretic perspective. 

In practice, there are experimental analogues or estimates of the Entropy Rate, such as the Sample Entropy and Approximate Entropy \cite{39}. We recently used Approximate Entropy to quantify dyadic social interaction in a clinical setting using data from wearable digital sensors and digitized the Autism Diagnostic Observation Schedule (ADOS, version2) \cite{40}. Therefore, we opt to use it in the current context as a proxy for measuring the information content of facial micro expressions.  

\subsection{A Unified Information Theoretic and Dynamical Pipeline for Facial Micro-expression Biometrics}

We design several parameter spaces to visualize the data according to the various metrics from nonlinear dynamical systems and information theory. We also include the average Fourier spectrum frequencies of the velocity data \cite{41} and represent the scatter of points in the parameter spaces colored ASD or NT, according to the clinical labels. Figure 3 provides the pipeline of these analysis that features parameter spaces with two representative participants. 

\subsection{Trigeminal Division Feature Averaging}

Upon partitioning facial landmarks into three distinct regions related to the trigeminal nerve divisions \cite{42}, 

\begin{itemize}
\item Opthalmic division (V1 region)
\item Maxillary division (V2 region)
\item Mandibular division (V3 region)
\end{itemize}

we averaged the landmark features for each of these regions. This offers a biologically compatible and compact characterization of facial dynamics. Due to the high correlations between adjacent facial regions, this approach also serves to compress information and to develop classifiers from a limited sample size, while distinguishing the motor output of the three distinct trigeminal nerve subdivisions.

\subsection{Quantifying Departure from Resting State at a Cohort Level}

The average features obtained for each subject have a certain distribution across NT or ASD populations, which may vary between different micro-expressions. As mentioned earlier, we reject any assumptions of normality regarding the data; we estimate confidence intervals using bootstrapping (see mathworks \cite{43} and links to the literature explaining the methods here \cite{44}). 

Thus, we compute empirical histograms for the three micro-expression states and quantify the distance between the empirical distributions at rest vs. the other states. We can do this nonparametrically, using the Wasserstein distance (also known as the Earth Mover's distance \cite{45}). This is a proper metric in the space of probability distributions \cite{46}. 

\section{RESULTS}

\subsection{Dynamical Features Shift Differently Between ASD and NT participants from Resting to Active Facial States}

The parameter space spanned by each of the metrics of interest along the V1, V2 and V3 dimensions provides a way to visualize the scatter for each of the facial states. Figure 4 shows a subset of 4 features considered for the smile condition. These include the Lyapunov Exponent, Synergies, Average Frequency and micropeaks Spike Rate. They show computed region averages for 44 NTs, 73 HS ASD and 20 LS ASD. We also computed such plots during resting and surprise tasks. which can be found in the supplementary material. We can see trends emerging, i.e., larger Lyapunov exponent and synergies values, lower rate values, and a wider range of speed frequency averages for ASD, particularly those with HS needs. 

Lyapunov exponents measure the rate of separation of two trajectories arbitrarily close to each other. Higher values in ASD imply that their speed trajectories are more sensitive to small changes or perturbations than their NT counterparts.This renders the trajectories less predictable to an observer. It also renders them less predictable to the person himself, due to the closed-loop nature of the facial motor activity. The facial motor activity typically serves as feedback to the person, and the flow e.g., in conversation, is controllable. Yet, if the person's facial motor code is highly uncertain and unpredictable, the feedback from it will be rather deleterious to the voluntary control of emotions. A disconnect between the intended emotion and the involuntary projection of it will confuse the receiver of this information at the other social end of the communication channel. Likewise, more facial synergies in the ASD participants relative to NT controls hint at fewer DOF in ASD. Abundant DOFs on demand are important to achieve flexible and adaptable voluntary control of self-generated micro expressions. This voluntary control on demand would be important to dynamically master desired emotions at will, in flight, overcoming involuntary motions during emotional exchange and social communication attempts. 

\begin{figure}[ht]
\centering
\includegraphics[width=0.4\textwidth]{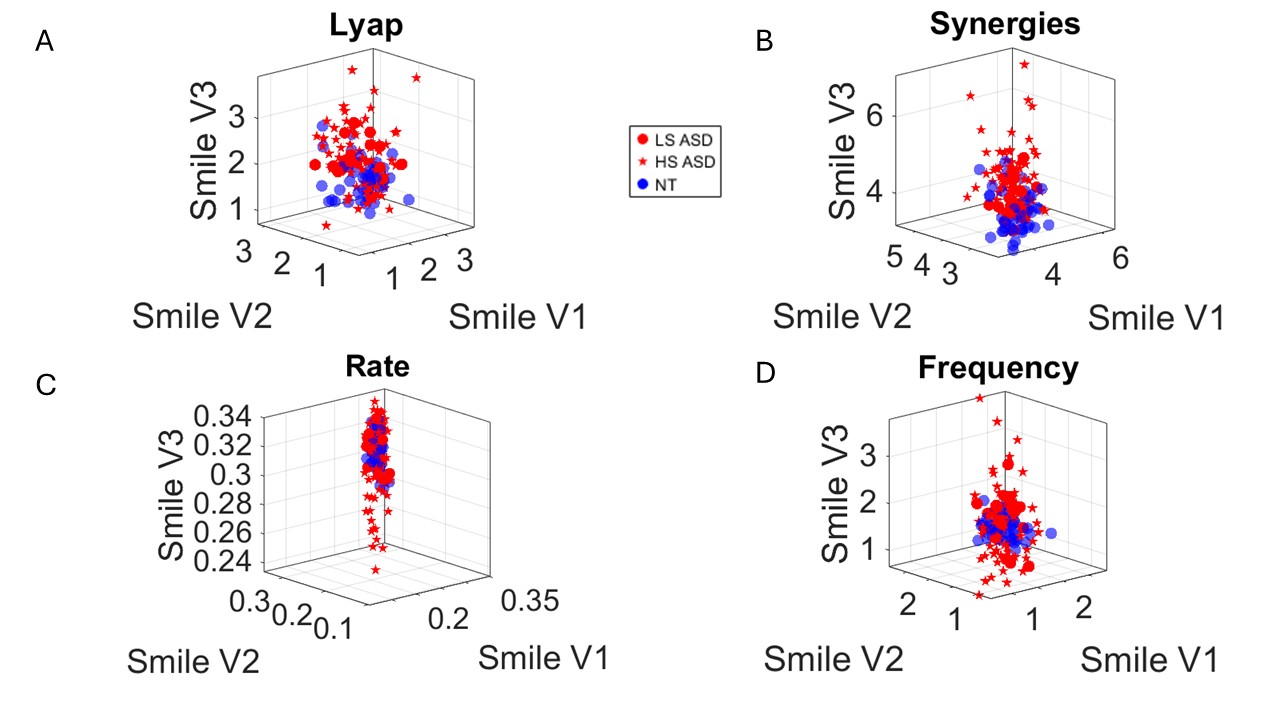}
\caption{Parameter spaces spanned by facial region averages, during the smile micro-expression task. (A) Scatters comprise NT, HS ASD and LS ASD Lyapunov exponent. (B) Synnergies obtained from the geometric approach involving SPD covariance matrices. (C) Average rate (number of micropeaks peaks per unit time). (D) Mean speed frequency values.}
\label{fig:Panel 4}
\end{figure}

If $A_{1},A_{2},A_{3}$ are the values of a feature for the $V_1,V_2,V_3$ areas, we can compute the quantity $\sqrt{A_{1}^{2} + A_{2}^2 + A_{3}^2}$, which we call the "norm" of the feature. Then, for each feature, we can derive empirical histograms of that norm for the ASD and the NT cohorts, during the resting state, the smile, and the surprise micro expressions. Then, we quantify the "departure" from resting state to smile or surprise state as the Earth Mover's Distance (EMD) between the empirical histogram at the resting state and the empirical histograms at the active state, respectively.

As with other biorhythmic movement data across the body, here we find that the continuous Gamma family of probability distribution fit the facial data in an MLE sense. Figure 5 shows the results of this analysis. A consistent pattern emerges across these features, except for the embedding dimension. As the cohort shifts from resting state to some active state (in this case smile), the probability distribution for the NT population becomes more concentrated and the overall spread decreases. However, the ASD population fluctuates around these narrower NT ranges. Figure 5C shows the differences in the distributions between NT and HS ASD, LS ASD quantified by the EMD. This panel reveals which parameters most clearly departure from NT levels. It also highlights self-emerging differences between the ASD subtypes with HS ASD consistently separating from LS ASD.

\begin{figure}[ht]
\centering
\includegraphics[width=0.4\textwidth]{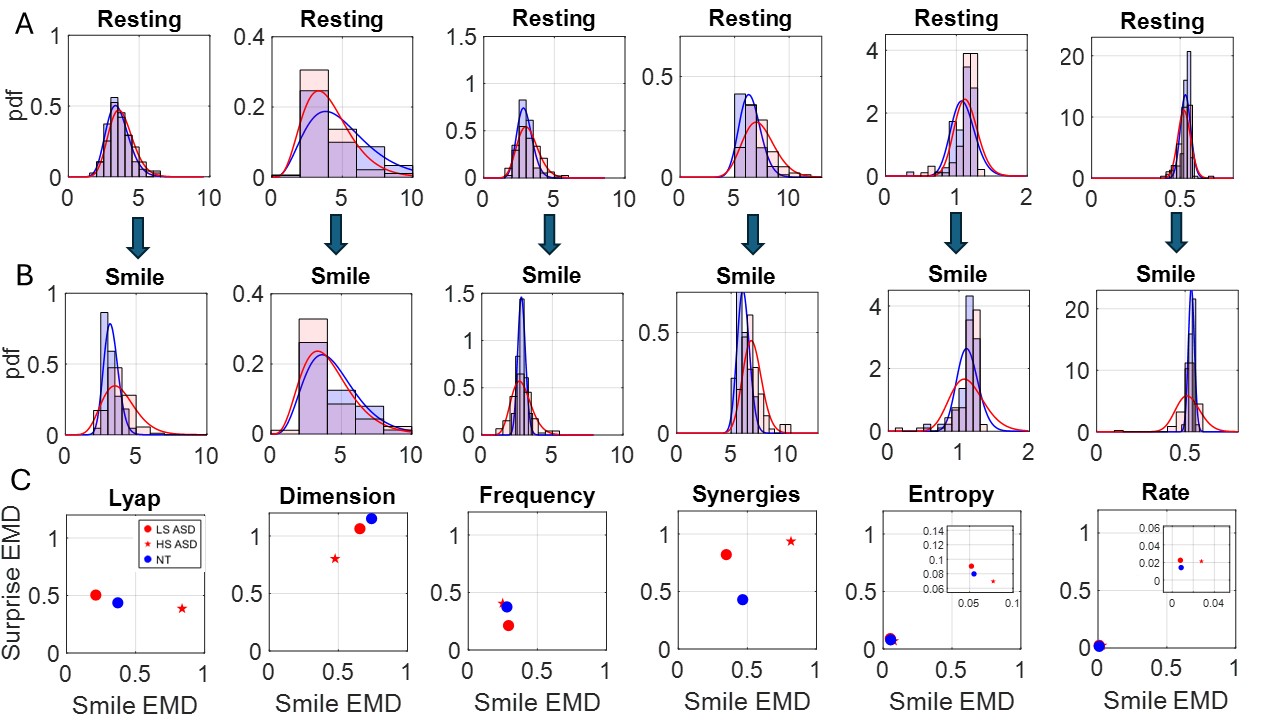}
\caption{Parameter distribution shifts in population statistics obtained from facial active states relative to rest. (A) Empirical histograms for the norm $\sqrt{A_{V_1} + A_{V_2} + A_{V_3}}$ of the average features of the three trigeminal division areas during resting state and the best fitting Gamma distribution in an MLE sense. NT and ASD are supperimposed. (B) Shifts in parameter distribution during the smile task. (C) Parameter plane featuring the Earth Mover's Distance values taken between population empirical histograms during surprise or smile micro-expression relative to resting state.Insets in Entropy and Rate parameters zoom in to show the patterns}
\label{fig:Panel 5}
\end{figure}

\subsection{Temporal Dynamics are Differentially Modulated by Noise-to-Signal Ratio in ASD vs. NT}

The micropeaks (MMS) framework that assesses the noise-to-signal ratio of speed in bodily motions has recently been extended to the face domain to characterize hidden affective features in ASD \cite{7}. We ask if the present geometric, nonlinear dynamical and information theoretic features introduced in this study correlate with the noise-to-signal ratio of the facial speed that our recent work identified.

In Figure 6, one can appreciate the parameter plane spanned by the norms of four parameters that we identified in Figure 4 as promising, as a function of the noise-to-signal ratio. Along each parameter (column-wise A,B,C) we appreciate the shift of the scatter across different parameters, hinting at a rich source of information to possibly distinguish different features of the NT vs. ASD subtypes. Focusing on the NT, we see how they change from resting state to active motions (smile and surprised micro expressions). In addition, we see the spread of the ASD subtypes, with far broader ranges in the HS ASD group that consistently departs from both the NT and the LS ASD.

\begin{figure}[ht]
\centering
\includegraphics[width=0.4\textwidth]{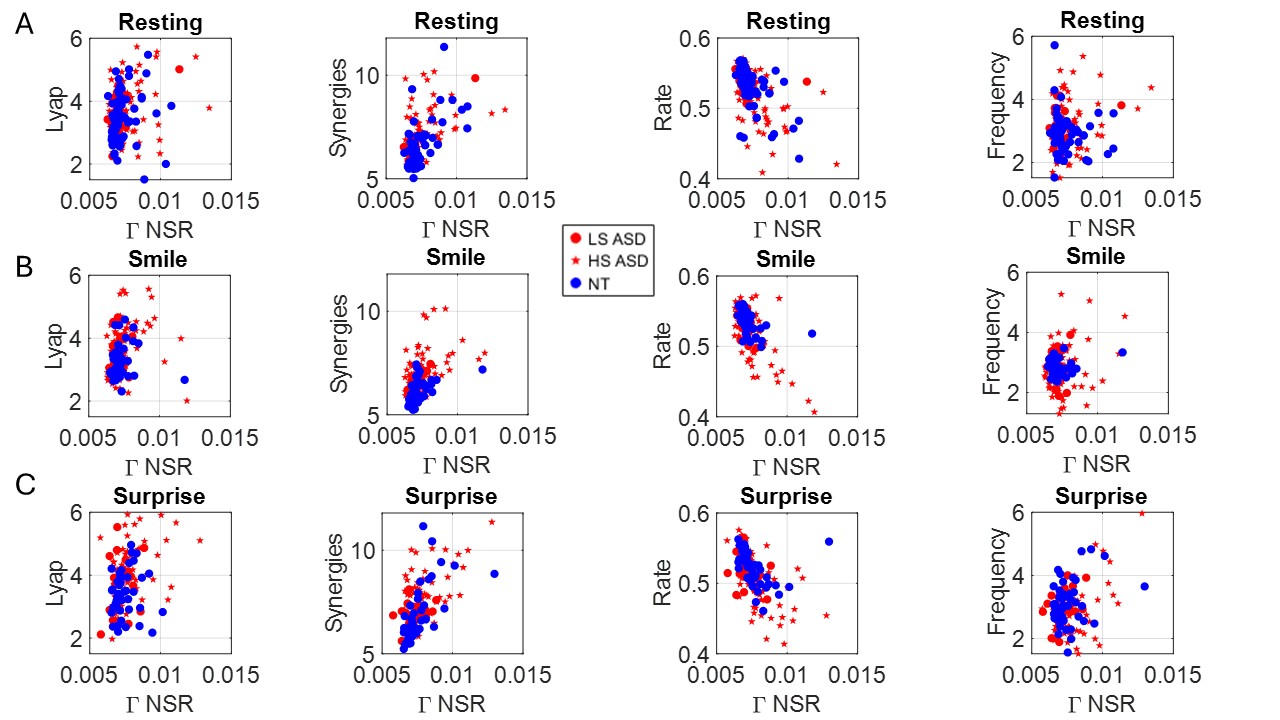}
\caption{Scatter shifts on the parameter plane spanned by the $\Gamma$ NSR vs. parameter identified in Figure 4 as revealing maximal differences in smile and surprise relative to resting state. For visualization purposes, noise coordinates are restricted to the range $[0.005 0.015]$, appendix plots include the few outliers not appearing in these plots. (A) Resting state shifts across  each parameter as a function of the $\Gamma$ NSR. (B) Smile task. (C) Surprise task.}
\label{fig:Panel 6}
\end{figure}

The results relating the parameters to the $\Gamma$ NSR motivate us to properly quantify the relationship between the $\Gamma$ NSR ratio and the six remaining features. We fit multivariate linear relationships for the averages of each facial area $V_{i}, i=1,2,3$ and for all subjects and micro-expression tasks. The linear model has the following form, where $i$ refers to the facial region and $feat$ refers to the feature parameters under consideration:

\begin{equation}
A_{NSR}^{V_{i}} \sim  \sum_{feat=1}^{6} c_{feat}^{V_{i}} A_{feat}^{V_{i}}
\end{equation}

\begin{figure}[ht]
\centering
\includegraphics[width=0.4\textwidth]{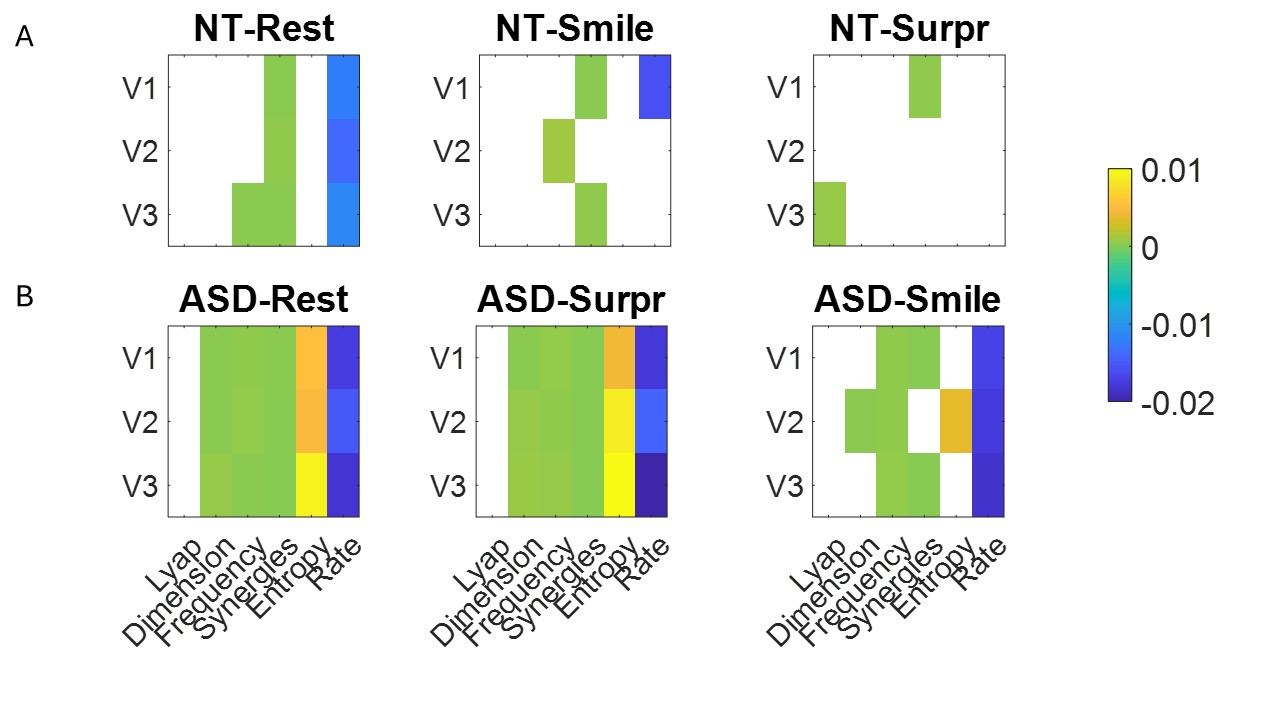}
\caption{Multivariate regression results between average trigeminal region features and $\Gamma$ NSR for NT and ASD groups for resting, smile and surprise micro-expression states. (A) NT cohort with color coded regression coefficients shown between each feature and $\Gamma$ NSR from the multivariate model whenever the coefficient value is significant ($p < .05$). (B) Results for the ASD cohort using the same representation. }
\label{fig:Panel 7}
\end{figure}

In Figure 7, we report the estimated correlation coefficients $c_{feat}^{V_{i}}$, whenever they are statistically significant ($p < .01$). The correlations observed in Figure 6 are now confirmed. Positive correlation coefficients between frequency, synergies, and $\Gamma$ NSR and negative correlation coefficients between rate and $\Gamma$ NSR.  Switching to smile and surprise states, the correlation is non-significant overall. In the ASD cohort, the correlations differ from the NT in panel A. Consistent and significant negative correlations between rate and $\Gamma$ NSR across all micro expressions are observed along with positive correlations between all other features (except Lyapunov exponents) and $\Gamma$ NSR, particularly between entropy and the $\Gamma$ NSR.  Positive correlations become less pronounced under the smile micro-expression state. We conclude that peak activity away from the empirical mean (rate) is consistently negatively modulated by the $\Gamma$ NSR across the ASD group. Also, from our previous empirical work, we know that moment-to-moment variability, as measured by the $\Gamma$ NSR, implies unpredictability in the motor code (approximate entropy) due to its tight (negative) scaling relationship with the Gamma shape. As the $\Gamma$ NSR increases, the Gamma shape decreases towards the memoryless exponential regime of the Gamma family \cite{3}, denoting greater randomness. In high random $\Gamma$ NSR regimes, the predictability of the motor code decreases.

In the ASD cohort, the rate is negatively correlated with the $\Gamma$ NSR, but all other parameters, especially approximate entropy, are positively correlated. This noise-feature modulation phenomenon appears to be much less pronounced in the NT population, a result that is consistent with previous findings from our lab \cite{3}, \cite{47}. 

\subsection{Tracking Parameters' Norms Across Different micro expressions Separates ASD from NT}

Marked differences across the cohorts are reported in Figure 8A where we show population averages for the norm parameter values in the LS ASD, HS ASD and NT groups. These are obtained for each activity and at resting state. In addition, we empirically estimate confidence intervals using bootstrapping. 

The averages of the Lyapunov exponents, facial synergies, and approximate entropy are trending higher in the ASD cohort than in the NT cohort. More importantly, this is consistent across micro expressions. The results agree with those obtained in Figure 4.  

The embedding dimension is on average larger for the neurotypical group, for all three tasks, that is, the phase-space of facial landmark velocities has more dimensions in the neurotypical group. This implies that in autism, the space of possible facial trajectories has "fewer degrees of freedom" (DOF), which confirms the hypothesis that we drew from Results Sections A and B. 

\subsection{Classifier Separates Training States}

We used all parameter norms obtained for the (V1,V2,V3) points in each of the parameter spaces, including the $\Gamma$ NSR ratio, as predictors in an ensemble classifier model (specifically subspace k-nn), available in Matlab Statistics and Machine Learning toolbox. For more information on ensemble classification, see \cite{48}. We used 10 folds for cross validations. 

In Figure 8B, the reader can appreciate the classification results in which $77.9$ percent of the ASD cohort and $72.0$ percent of the NT cohort were correctly classified. Given the balanced proportion of ASD to NT in the sample size, the model did not under-diagnose or over-diagnose. The overall validation accuracy is $75.3$ percent, despite the overall modest size of our cohort. 

\begin{figure}[ht]
\centering
\includegraphics[width=0.4\textwidth]{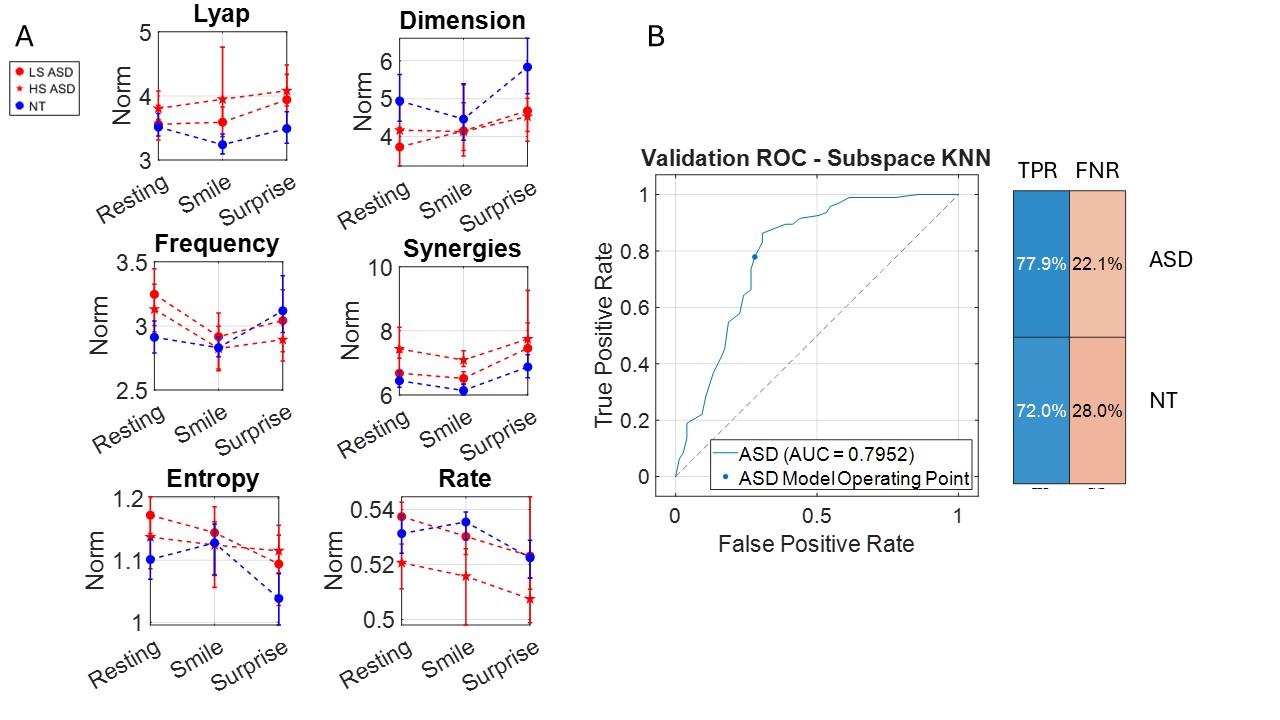}
\caption{Parameter differentiation across tasks and NT vs. ASD automatic classification. (A) Six parameters's means empirically estimated and evaluated across tasks and cohorts plotted with confidence intervals from bootstrapping. (B) Results from subspace KNN classifier trained on feature norm (10 cross validation folds) outperformed all other classifiers with a validation accuracy of 75.3 }
\label{fig:Panel 8}
\end{figure}

\section{CONCLUSIONS}

This work aimed at maximizing the use of information contained in nuances of our facial behaviors. We did so by examining the geometry of facial grid points' covariations, the nonlinear dynamics of the facial micromotions and the information content of facial microgestures, as they transitioned from resting to more active states. We posited that trapped in such nuances, often discarded as gross data, was a wealth of information that could be used to measure the departure of atypical neurodevelopment from the typical developmental course. To that end, we derived several parameter spaces and empirically characterized ranges of parameters that rendered them useful in automatic classification. The overall results of this work revealed different parameter landscapes that could be integrated to both facilitate interpretation of the analyses and offer a brief way to quantify facial micro expressions.

The empirical estimation of parameters characterizing micro expressions in participants with ASD of variable supporting needs and in NT participants was done under two main levels of inquiry: A micro-level of inquiry involved speed micropeaks derived from the rates of change in pixel positions over the span of 5 seconds / assay, while a macro-level of inquiry referred to muscle action units activations. We employed tools from differential (Riemannian) geometry to assert synergies across facial subregions involved in emotional and social exchange. Besides the geometric analysis, we borrowed several tools from nonlinear dynamical systems and information theory to further characterize several features of facial micromotions.  

Despite the brevity of our video acquisition assays, we found that useful signals in the data are suitable to differentiate participants in our cohorts. Exploring different classifiers, we identified the optimal one offering an accuracy of 75.3 percent for very brief data summarized by the parameters that our methods identified. Most importantly, this was achieved using empirical estimates of parameters of continuous probability distribution functions that are useful to apply in generative models of AI. A handful of features from just 5 second-long assays capturing facial micro expressions in a modest cohort of 124 subjects separated participants in the broadly heterogeneous ASD group from those in the NT group. Due to the brevity of our assays and the effectiveness of these parameter identifications, we are confident that more frequent measures can be obtained in a longitudinal study setting.

Each of the features identified by our methods is interpretable and reveals information regarding predictability, stability of facial trajectories, synergistic phenomena across distinct facial areas, information in the temporal code, $\Gamma$ NSR ratio of standardized facial micropeaks, and rate of peak speed activity away from average activity for localized landmarks. The reason for computing those features was inspired by the structure of the trigeminal nerves, the three bundles of nerves providing afferent input relevant to the control of ophtalmic functions (eye motions, gaze, saccades, eye fixation, etc.), maxillary functions (nose micromotions related to odor sampling, sneezing, and general memory formation and retrieval, among others), and mandibular functions important in speaking, mastication and even reflecting patterns related to swallowing and the avoidance of choking. Although our assays were brief and simple, the biometrics that we have derived in this work can be extended to other assays to assess speech, stroke rehabilitation, and other functions in different disorders of the nervous systems. 

Methods from different active areas of research revealed the relationships in the various correlations that we reported. Among these are correlations between the $\Gamma$ NSR ratio from stochastic analyzes and the temporal facial properties of the dynamical systems framework. The rate of micropeaks activity, that is, the number of speed fluctuations away from the empirical mean per unit time, generally decreased as the $\Gamma$ NSR increased,
the effect being particularly prevalent and consistent across facial regions, in the ASD group. Furthermore, stochastic characteristics and information-theoretic metrics were related in the positive correlations found between the $\Gamma$ NSR and the entropy rate. The ASD subpopulation also had on average fewer dimensions in the reconstructed phase-space but more synergies across local face regions. 

The positive relationship between facial synergies and $\Gamma$ NSR in autism could explain our previous hypothesis. The reduction of overall DOFs due to excessive facial synergies makes the control of facial expressions challenging, leading to excessive motor noise. This is also reflected in the correlation between the $\Gamma$ NSR and entropy, a quantity that measures temporal randomness, and in the rate measures reflecting the fluctuations in speed maxima away from the empirical mean. Therefore, from the inverse relationship between $\Gamma$ NSR and the rate in the autistic population, we conclude that excessive noise in autism reduces significant facial activity beyond the baseline state.  

These findings suggest that the underlying motor system that controls facial expressions has fewer DOF in ASD. This may be due to excessive involuntary synergies between neighboring muscle units. With fewer DOF emerging from involuntary synergies it is possible that the fluidity of transitioning from one micro-expression to another decreases, thus potentially impeding the ability to control, on demand and at will, the dynamics of facial micro expressions during social exchange. Given the closed-loop nature of emotional gestures, this restricted motor control could also impact the overall repertoire of variations in micro-gestures and help explain why facial dynamics in the ASD cohort were more unpredictable and chaotic. Another possibility, one that would explain the lower rate at higher levels of noise in autism, is the existence of some sort of compensatory mechanism, whereby noise decreases to compensate for the abundance of disruptive facial fluctuations. 

The present results significantly extend our current understanding of facial dynamics by integrating computer vision/ deep learning methods that extract landmark kinematics from active face videos, with methods from the modern theory of dynamical systems, information theory, and geometric theory of matrices. These analyzes were performed using new analytical techniques informed by a large body of empirical knowledge acquired by our lab. They begin to provide ground truth for artificial generative methods. Our previous work characterizing the stochastic features of human biorhythmic bodily activities under a new unifying statistical platform for individualized behavioral analysis is now extended to facial micromotions. 

The ease and brevity of our mobile assays, paired with the empirical characterizations and parametric estimation presented here, will enable the high scalability of our research and a diversification of phenotypic data to further estimate parameter ranges across humans. We believe that the present work will contribute to the further development of the field of affective computing.

 
\section*{Acknowledgment}
The authors thank the families who participated in the study and the Nancy Lurie Marks Family Foundation for funding this project. 
\ifCLASSOPTIONcaptionsoff
  \newpage
\fi


%

%

\begin{IEEEbiography}[{\includegraphics[width=1.2in,height=1in,clip,keepaspectratio, angle=-90]{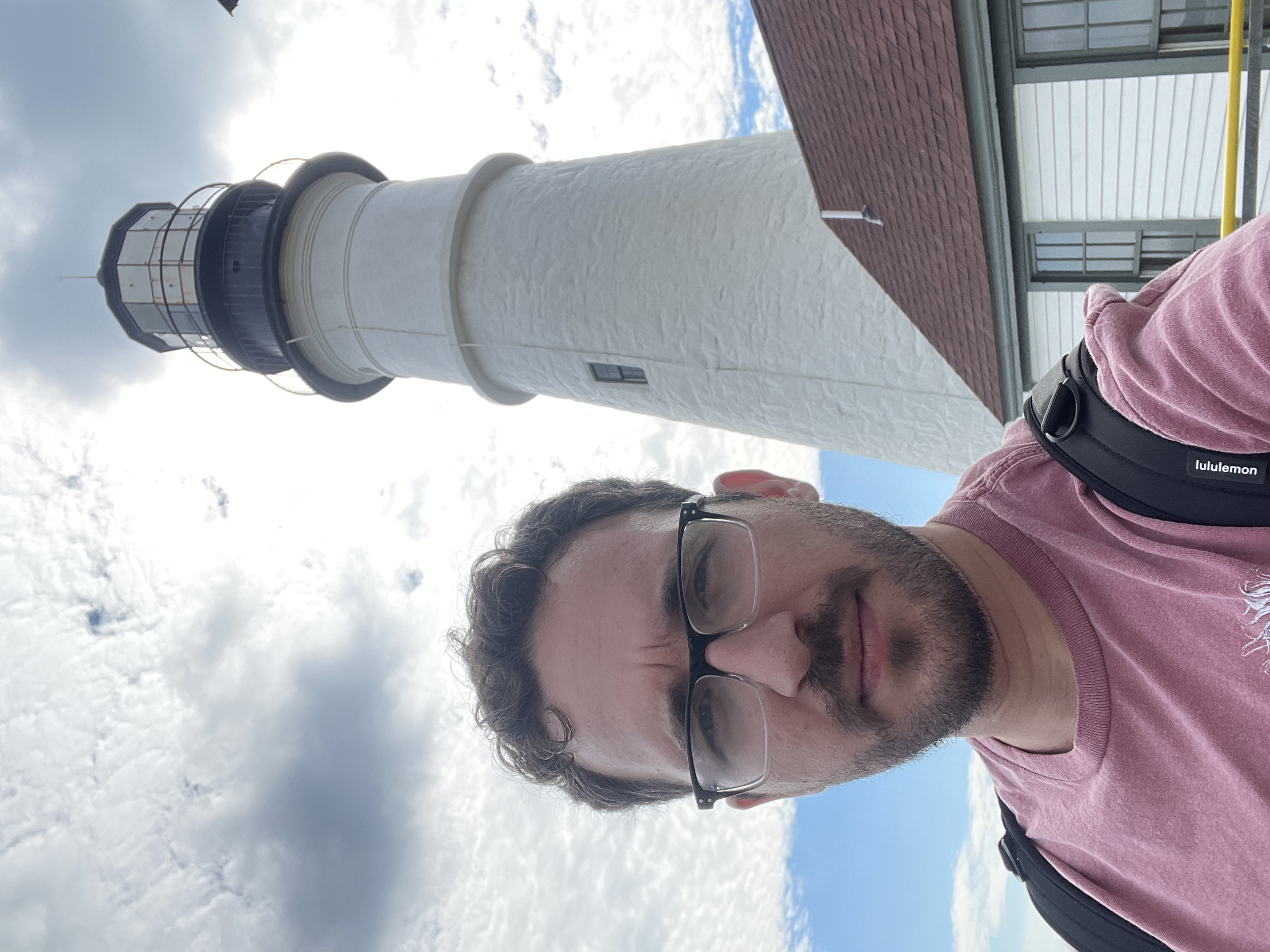}}]{Theodoros Bermperidis}
Dr. Theodoros Bermperidis is a postdoctoral fellow at the Sensory Motor Integration Laboratory of Rutgers University. He completed his Ph.D. in Cognitive Science with a focus on the development of new computational models of motor agency, with applications to ML and AI. He holds a BSc \& MSc  degree in Electrical \& Computer Engineering and Biomedical Engineering from the University of Patras, Greece, and a Master's degree in Cognitive Science with applications to digital health.
\end{IEEEbiography}
\begin{IEEEbiography}[{\includegraphics[width=1.1in,height=1.22in,clip, angle=0]{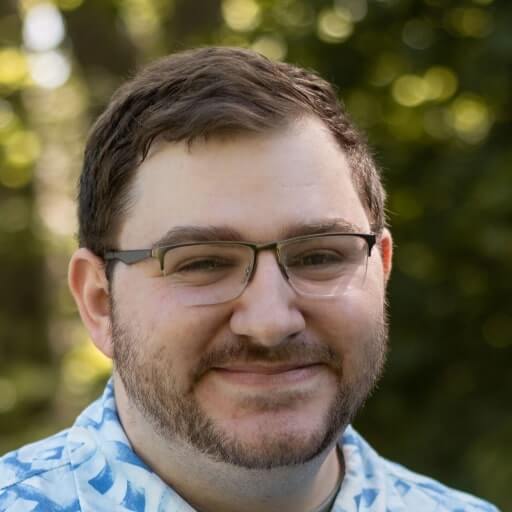}}]{Joe Vero} 
\
Mr. Joe Vero is a graduate student in the Sensory Motor Integration Laboratory of Rutgers University. He completed a Bachelor's degree in Biomedical Engineering and holds Master's degrees in Data Science and Cognitive Psychology. He is pursuing a Ph.D. in Cognitive Psychology with a focus on computational models of motor control and app development for digital health.
\end{IEEEbiography}

\begin{IEEEbiography}[{\includegraphics[width=1.3in,height=1in,clip,keepaspectratio, angle=0]{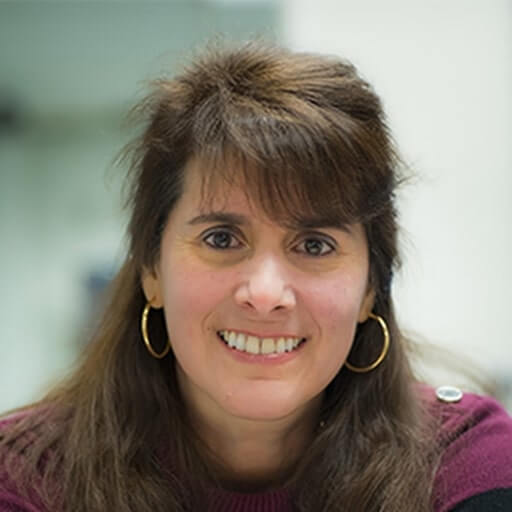}}]{Elizabeth B Torres} 
Dr. Elizabeth B Torres is a Professor at Rutgers the State University of New Jersey, USA, and the Director of the Sensory Motor Integration Laboratory of the Psychology Department. She is a member of the Rutgers University Center for Cognitive Science and of the Computer Science Department Center for Biomedicine Imaging and Modeling. She is a graduate faculty member of the Department of Biomedical Engineering and of the Neuroscience and Cell Biology Program. Torres holds a Bachelor's degree in Computer Science and Applied Math, a Ph.D. in Cognitive Science, and Postdoctoral training in Computational Neural Systems and Electrophysiology. She is a US inventor and innovator with multiple granted US patents related to the objective and automated analyzes of voluntary and spontaneous human behavior, with digital health applications spanning from newborns to the elderly.
\end{IEEEbiography}




\end{document}